\documentclass[
aps,
prl,
preprint,
groupedaddress,
reprint,
twocolumn,
floatfix,
superscriptaddress,
amsmath,amssymb,
]{revtex4-2}

\usepackage{graphicx}
\usepackage{dcolumn}
\usepackage{bm}
\usepackage[hyperindex,breaklinks]{hyperref}
\usepackage{amsmath}
\usepackage{graphicx}
\usepackage{CJK,braket}
\usepackage{mathrsfs}
\usepackage{bbm,color}
\usepackage{caption}
\usepackage{subfig}

\def\sdg{Schr\"odinger}
\def\td{\text{d}}
\def\ti{\text{i}}
\def\tg{g_c}

\begin{document}


\title{Microscope for Quantum Dynamics with Planck Cell Resolution}


\begin{CJK}{UTF8}{gbsn}
\author{Zhenduo Wang (王朕铎)}
\affiliation{International Center for Quantum Materials, School of Physics,
Peking University, 100871, Beijing, China}
\author{Jiajin Feng (冯嘉进)}
\affiliation{International Center for Quantum Materials, School of Physics,
Peking University, 100871, Beijing, China}
\author{Biao Wu(吴飙)}
\affiliation{International Center for Quantum Materials, School of Physics,
Peking University, 100871, Beijing, China}
\affiliation{Wilczek Quantum Center, School of Physics and Astronomy,
Shanghai Jiao Tong University, Shanghai 200240, China}
\affiliation{Collaborative Innovation Center of Quantum Matter, Beijing 100871,  China}


\date{\today}

\begin{abstract}

We introduce the out-of-time-correlation(OTOC) with the Planck cell resolution. The dependence of this OTOC 
on the initial state makes it function like a microscope,  allowing us to investigate the fine structure of quantum dynamics 
beyond the thermal state.  We find  an explicit relation of this  OTOC to the spreading of the wave function in the Hilbert space, 
unifying two branches of the study of quantum chaos: state evolution and operator dynamics.
By analyzing it in the vicinity of the classical limit,  we clarify the dependence of the OTOC's exponential growth  
on the classical Lyapunov exponent. 

\end{abstract}


\maketitle
\end{CJK}
{\it Introduction} -- In recent years, the out-of-time-order correlation (OTOC)~\cite{Larkin1969,Maldacena2016} has attracted 
great attention in the field of quantum dynamics~\cite{Hashimoto2017,Chen2018}. It offers us a powerful tool to quantify 
quantum chaotic behavior, in particular, in many-body systems~\cite{BAGRETS2017727,Bohrdt_2017,PhysRevD.94.106002}. 
However, there remain some fundamental issues,  two of which are what we try to resolve in this work.

\begin{figure}[t]
  \includegraphics[width=\columnwidth]{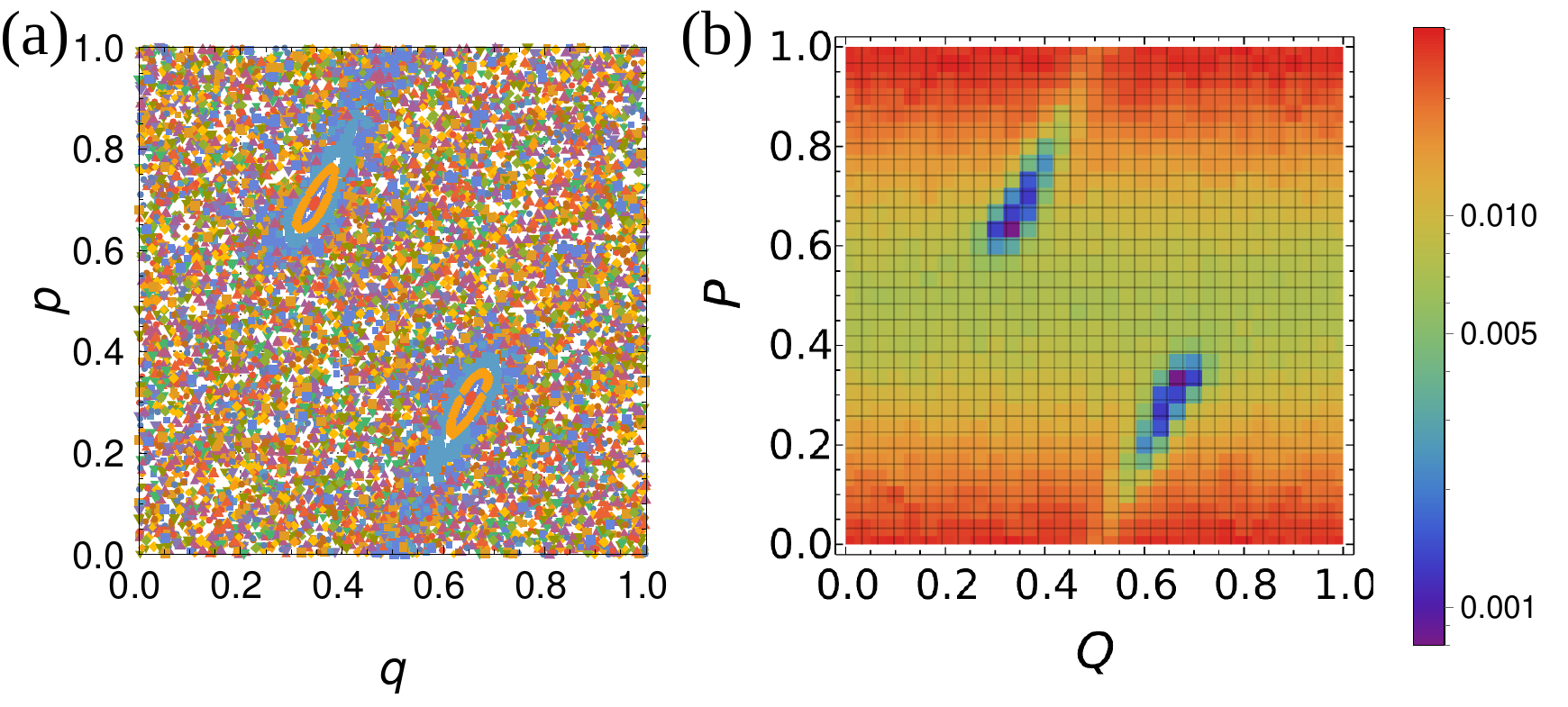}
  \caption{ (color online) The comparison between the classical Poincar\'e section and its quantum version for kicked rotor. 
    (a) The classical one is generated by random sampling in phase space and with $100$ kicks. (b) The 
    quantum version with effective Planck constant $\hbar\approx 0.007$ is constructed by computing 
    the OTOC defined in Eq. (\ref{eq:otoc}) for each Planck cell with $70$ kicks. The blue valleys of OTOC correspond to the classical integrable islands. The kicking strength is $K=4.7$.}
  \label{fig:krSec}
\end{figure}

The cause of  growth of OTOC is not clear yet. It has been found that the early-time growth of OTOC is related to the classical Lyapunov exponent~\cite{Garcia-Mata2018,Rozenbaum2017,Chavez-Carlos2019,Jalabert2018,Cotler2018}. This makes OTOC popular in the study of quantum chaos. However, recent works demonstrated such exponential growth can be caused by a saddle point but not chaos~\cite{Xu2020,Hashimoto2020}. One of the reasons is that the usual OTOC\cite{Maldacena2016}(thermal OTOC below) has no  
dependence on the initial conditions, which is a general feature of all dynamics. In particular, as is well known,  
the dynamics of the same classical system can be regular for one set of initial conditions and chaotic for another set of  
initial conditions. This is usually illustrated with the Poincar\'e section (e.g., see Fig.~\ref{fig:krSec}(a))~\cite{arnold2013mathematical,Lichtenberg1992,reichl2013transition}. 
Thus we cannot directly use the growth of OTOC as an indicator to classify dynamics.

Quantum chaotic behavior can also be characterized by the wave packet spreading~\cite{PhysRevLett.52.1,PhysRevLett.55.645,KORSCH1981627,Han2015,qmixing,Moudgalya2019,wang2020genuine}. 
This \sdg~picture dynamics corresponds to our intuitive understanding to chaotic motion. In contrast, 
the OTOC reflects the quantum dynamics in the Heisenberg picture~\cite{PhysRevB.99.094312,Nahum2018}.  
Some evidence implies that these two pictures are related~\cite{Chen2018, Moudgalya2019}, 
but an analytical derivation is still lacking.

To resolve the above issues, we introduce a modified version of OTOC:
\begin{equation}
  C(t, x) = -\bra{x}[\hat Q(t),\hat P(0)]^2\ket{x} \label{eq:otoc}\,.
\end{equation}
The state $\ket{x}$ for one dimension is
\begin{equation}
  \ket{x}\equiv \ket{Q_x,P_x} = \frac 1 {\sqrt{\Delta_q}}\int_{Q_x}^{Q_x+\Delta_q} \td q \ \ket{q}e^{\ti P_x q/\hbar}
  \label{eq:qpbasis}.
\end{equation}
The generalization to higher dimensions is straightforward. Variable $x$ traverses all Planck cells(squares in Fig. \ref{fig:krSec}(b)) by $x=(Q_x,P_x) = x_0+(m \Delta_q, n\Delta_p)$, where $m, n$ are integers, $\Delta_q, \Delta_p$ are size of cells along $q, p$ axes, and $x_0$ is the origin of phase space. These states $\ket{x}$ form a set of complete orthonormal basis[appendix.~\ref{phbsproperty}], and they are localized in both position and momentum space~\cite{Jiang2017,PhysRevE.100.052206,Fang2018}. Operators $\hat Q = \sum_x \ket{x}Q_x \bra{x}$ and $\hat P=\sum_x \ket{x}P_x\bra{x}$ are so-called macroscopic position and momentum operators~\cite{Neumann1929,vonNeumann2010,Han2015,Fang2018}.

There are already many variations of  OTOC~\cite{Hashimoto2017,LewisSwan2019,Maldacena2016,PhysRevLett.125.014101,Rozenbaum2017}.
Compared to these definitions and the original one~\cite{Larkin1969,Maldacena2016}, our definition of OTOC is state-dependent. 
That is, for different states $\ket{x}$, this OTOC has distinct long-time behavior.  As a result, we can use it to plot a quantum-version of
the Poincar\'e section. An example of quantum kicked rotor is shown in Fig. \ref{fig:krSec}(b), which captures the salient features 
in the corresponding classical Poincar\'e section in Fig. \ref{fig:krSec}(a). Since the thermal OTOC\cite{Maldacena2016} averages 
over all quantum states,  our OTOC functions like a microscope for quantum dynamics with Planck cell resolution.  
Furthermore, we can analytically show 
that the dynamical behavior of this OTOC is explicitly related to the wave packet spreading and clarify the
dependence of the OTOC growth on the classical Lyapunov exponent.

{\it OTOC and wave packet spreading} -- 
The long-time behavior  of OTOC in Eq. (\ref{eq:otoc}) can be shown  related to   the  wave packet spreading explicitly. 
To see this, we consider 
the semiclassical limit $\hbar\rightarrow 0$. At this limit, the matrix elements of the propagator $\bra{q'}\hat U\ket{q} \sim e^{\ti S(q,q')/\hbar}$\cite{sakurai1985modern} in the basis of Planck cells can be written as
\begin{equation}
  \begin{split}
    \bra{Q',P'}\hat U\ket{Q,P} \propto&\int_{Q'}^{Q'+\Delta_q}\td q' \int_Q^{Q+\Delta_q} \td q \\
    &\exp\Big(\ti \frac {Pq-P'q'+S(q,q')} {\hbar}\Big)\,.
  \end{split}
  \label{eq:propa}
\end{equation}
Because of the stationary-phase approximation\cite{bleistein1986asymptotic}, the non-vanishing zeroth order term of above integral 
is determined by equations $P+\partial S(Q,Q') /\partial Q =-P'+\partial S(Q,Q')/\partial Q'=0$, which is the classical 
trajectory governed by action $S$\cite{arnold2013mathematical}. This suggests us to  rewrite Eq. (\ref{eq:propa}) as follows,
\begin{equation}
  \bra{x'}\hat U(t) \ket{x} = e^{\ti \phi(x)}\delta_{x',\tg x} + f(x', x,t)\label{eq:expan} ,
\end{equation}
where $x, x'$ are Planck cells as in Eq. (\ref{eq:qpbasis}) and $\tg x$ is the cell containing $gx$, which is the state driven classically beginning from $x$ at time $t$. The function $f$, which we call quantum spreading function, describes the pure quantum dynamics on  top of the classical dynamics. 
When $\hbar$ gets to zero, the function $f$ vanishes and Planck cells becomes continuous
such that $\tg  = g$.  In this limit, the pure quantum dynamics is completely suppressed and we are left with only the classical dynamics.

These two terms in Eq. (\ref{eq:expan}) contribute differently to the dynamics. The first term transports dynamically 
state $x$ into its classical target $gx$. 
Due to the finite size of Planck cells, $\tg$ implements a coarse-grained version of classical dynamics.
When the evolution time is shorter or around Ehrenfest time~\cite{Shepelyanskii1981,PhysRevA.65.042113}, this term dominates.
The second term breaks this classical picture and depicts how widely the wave packet spreads due to quantum effects. 
The function $f(x', x,t)$ begins to be significant after the Ehrenfest time and becomes dominate beyond another time scale called quantum time~\cite{Zhao2019}. 
In the following,  we show how they  contribute separately to OTOC and result in quantum chaos.

With Eq. (\ref{eq:expan}) we can show an explicit relation between the growth of OTOC and the wave packet diffusion. 
The leading order of OTOC in Eq. (\ref{eq:otoc}) is the second order of $f$ and can be written as (see Appendix \ref{mdOTOC} for derivation details)
\begin{equation}
  C(t, x) = \sum_{z} (P_z-P_x)^2(Q_{\tg z}-Q_{\tg x})^2\big|f(\tg z,x,t)\big|^2
  \label{eq:expect},
\end{equation}
where $(Q_x,P_x)$ are the coordinate and momentum of the Planck cell $\ket{x}$. 
The time-dependent terms are $(Q_{\tg z}-Q_{\tg x})^2$ and $|f(\tg z,x,t)|^2$. 
The former is the partial distance between trajectories and it can reflect  
the sensibility of the coarse-grained classical dynamics $\tg$ to the initial condition $x$ when 
$z$ is close to $x$. The latter one weights these terms in this summation. The region around $x$  in which $|f(\tg z, x,t)|^2$ is significant 
shows how widely the wave packet spreads. 
Previous works\cite{Chen2018, Moudgalya2019} use OTOC to measure the operator spreading by the unitary time evolution in quantum mechanics and find the influence of classical dynamics on the matrix elements of operators. Eq.~(\ref{eq:expect})  shows that the growth of OTOC is caused by the exploration of the wavefunction in Hilbert space and the dynamical sensibility of classical counterpart. 

\begin{figure}[t]
  \includegraphics[width=\columnwidth]{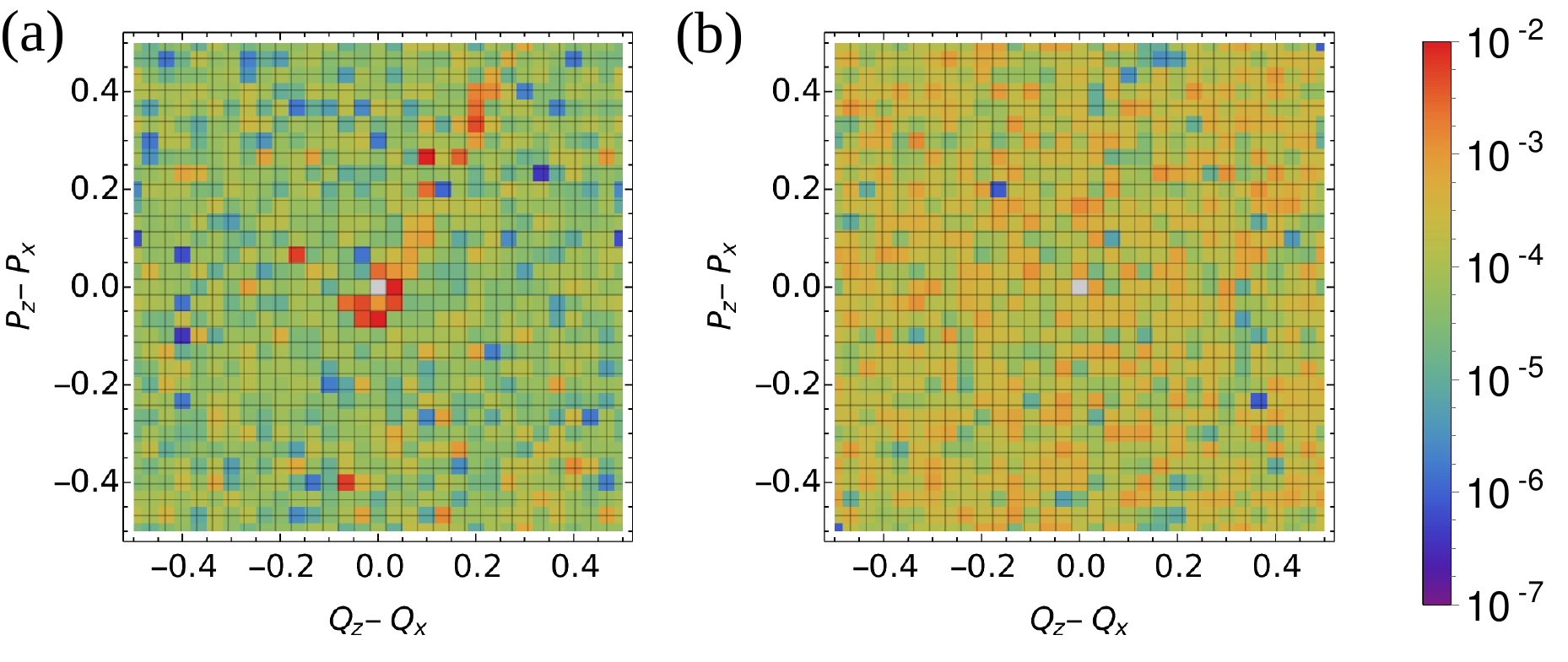}
   \includegraphics[width=\columnwidth]{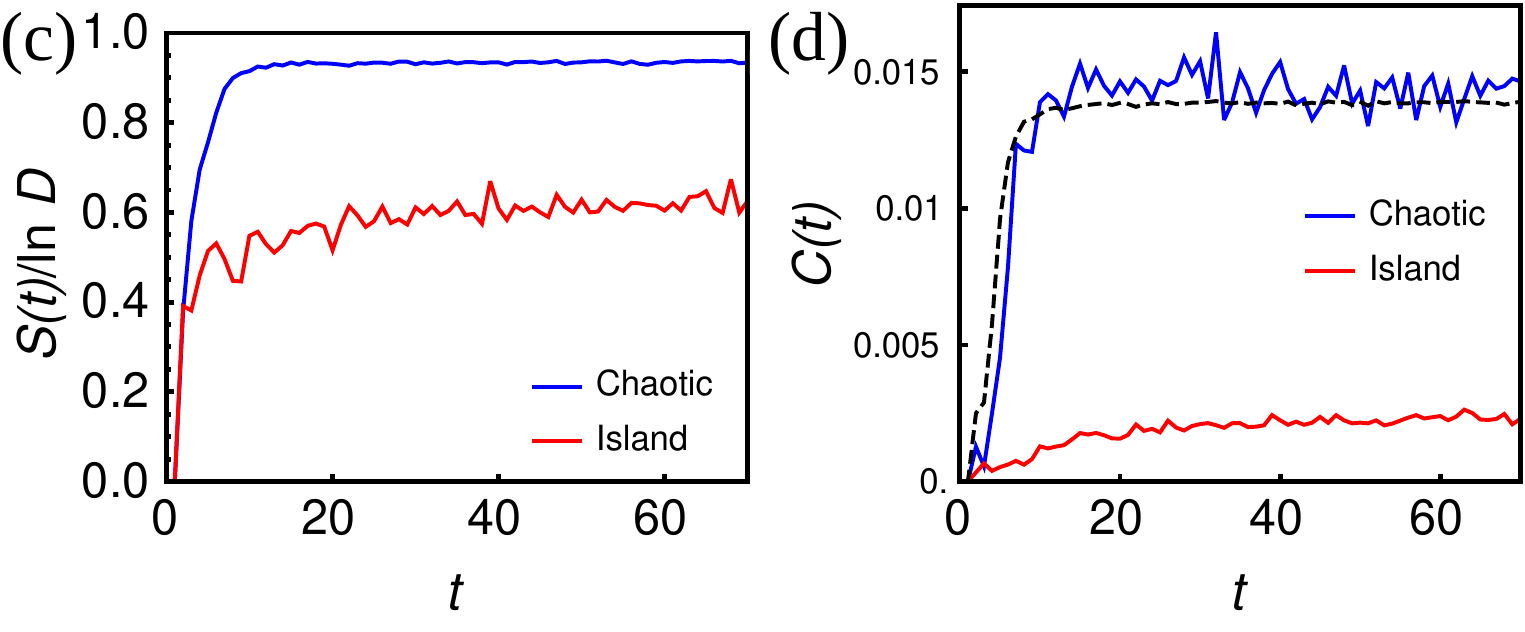}
  \caption{ (color online) The value of $|f(\tg z,x,t)|^2$ at points $z-x = (Q_z-Q_x,P_z-P_x)$ for the kicked rotor of kicking strength $K=4.7$. The number of
   kicks is $40$, and the effective Planck constant $\hbar\approx 0.007$. 
   The initial state for (a) is the Planck cell $x=(0.35,0.7)$ (inside an integrable island); the initial state for (b) is 
   $x=(0.2,0.2)$ (inside the chaotic sea). They are also, respectively,  the initial states for the ``Island" results and ``Chaotic" in (c,d).  
    (c) The growth of normalized GWvN entropy~\cite{PhysRevE.99.052117}. 
    (d) The growth of OTOC. The dashed line is the thermal OTOC at temperature 
    $T=\infty$ as in Eq. (\ref{eq:thermalotoc}). }
  \label{fig:qw}
\end{figure}

The features of the quantum  Poincar\'e section shown in Fig. \ref{fig:krSec}(b) are due to the function $|f(\tg z,x,t)|^2$. 
For $x$ initially in a classical integrable island,  $|f(\tg z,x,t)|^2$ can not spread as widely as it is in chaotic sea as clearly 
demonstrated by the numerical results in Fig. \ref{fig:qw}(a)(b). For the point $x$ located in integrable island, function $|f(\tg z,x,t)|^2$ is significant only when $z$ is close to $x$. This together with the regular motion in integrable island makes OTOC $C(t,x)$ saturate at small values. In contrast, for the point $x$ located in the chaotic sea, $|f(\tg z,x,t)|^2$ widely spreads over the chaotic sea and renders larger saturation values for the OTOC. In addition, we use the generalized Wigner-von Neumann entropy~\cite{PhysRevE.99.052117} to show how widely the wave packet spreads in the long run in Fig. \ref{fig:qw}(c). In Fig. \ref{fig:qw}(d), our OTOC behaves similarly. This further confirms that our OTOC is related to the wave packet spreading.

Thermal OTOC cannot illustrate the state dependency of dynamics like the quantum Poincar\'e section in Fig. \ref{fig:krSec}(b).
It computes the expectation value for  the thermal state at infinitely high temperature as
\begin{equation}
  \overline{C}(t) = -\frac{\text{Tr}[\hat U^\dagger \hat Q \hat U, \hat P]^2} {\text{Tr} \mathbbm{1}} = \frac 1 D\sum_x C(t,x) \label{eq:thermalotoc},
\end{equation}
where $D = \text{Tr}\mathbbm{1}$ is the dimension of the Hilbert space, and is also  the number of Planck cells. 
It is the average of $C(t,x)$ over the whole phase space or all the Planck cells with uniform weight. 

\begin{figure}[t]
  \includegraphics[width=\columnwidth]{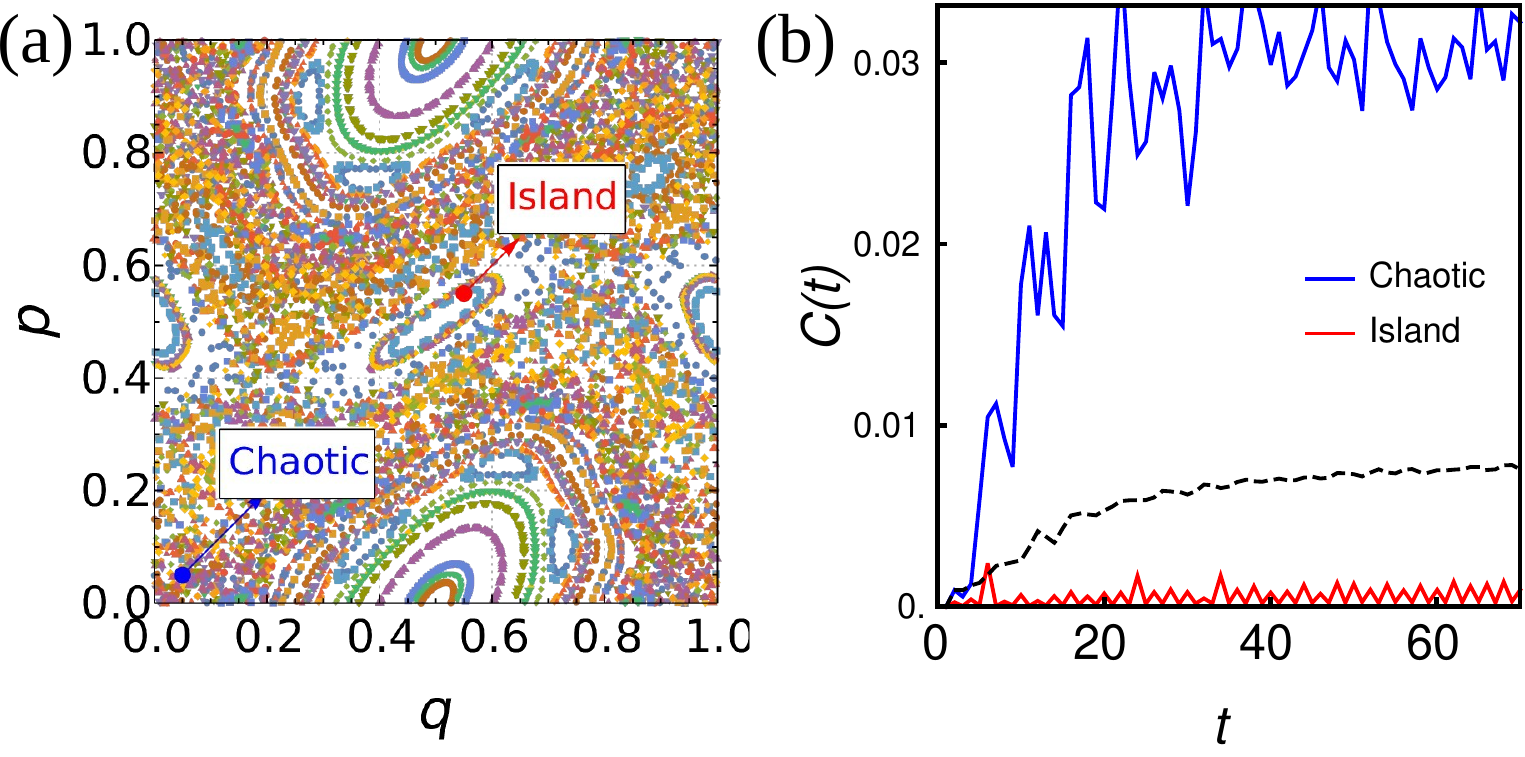}
  \caption{(color online) (a) The Poincar\'e section of classical kicked rotor of kicking strength $K=1.3$. (b) The time evolution 
  of our OTOC for two  Planck cell states corresponding to the labeled points in (a). 
  For comparison,  the thermal OTOC at temperature $T=\infty$ is also computed (dashed line). }
  \label{fig:krDyna}
\end{figure}

The behavior of $\overline{C}(t)$ is determined by the sizes of the areas occupied by integrable islands and chaotic seas in phase space. In Fig.~\ref{fig:qw} (d),  the thermal OTOC behaves like chaotic case (blue line) because the chaotic sea dominates. As a comparison, we consider a kicking strength that is close to $K_c \sim 0.972$ \cite{ChirikovScholarpedia}. The areas of integrable islands and chaotic sea are balanced as shown in Fig. \ref{fig:krDyna} (a). This leads to the dynamics in Fig. \ref{fig:krDyna} (b) in which $\overline{C}(t)$ is 
different from $C(t, x)$ for both integrable and chaotic cases.
If we regard OTOC as a microscope imaging the dynamical structure in the phase space, 
our OTOC in Eq. (\ref{eq:otoc}) has Planck cell resolution, which is the best allowed by quantum mechanics,  while the thermal one, namely $\overline{C}(t)$, can only render a smeared image with the crudest resolution 
that can only reveal whether the dominant part is integrable islands or chaotic sea.

{\it OTOC and Lyapunov exponent} -- 
The early-time growth of our OTOC is related to  the classical Lyapunov exponent. The quantum spreading 
function $f(x',x,t)$ is very narrow  during the early time evolution no matter whether $x$ is located in the integrable islands or chaotic sea. As a result, we are allowed to take $f$ out from the summation in Eq. (\ref{eq:expect}) and obtain
\begin{equation}
  C(t,x) \approx A\sum_{z } (Q_{\tg z} - Q_{\tg x})^2 \sim e^{2\lambda_Q t} , \label{eq:approtoc}
\end{equation}
where $A = \overline{(P_z-P_x)^2|f(\tg z,x)|^2}$ is a positive number and the summation is over the neighborhood of $x$.  The last term with $\lambda_Q$ is due to that $ (Q_{\tg z} - Q_{\tg x})^2$ is the result of the coarse-grained classical 
dynamics.  At the limit of $\hbar \rightarrow 0$,  the Lyapunov exponent of OTOC $\lambda_Q$ becomes 
the classical Lyapunov exponent $\lambda_C$ because of $\lim_{\hbar \rightarrow 0}\tg = g$. When time $t$ is not short and/or the Planck constant $\hbar$ is not small enough, the factor $|f|^2$ can not be taken out of the summation. Therefore, the exponential growth of OTOC is not
guaranteed in general because the evolution of $|f|^2$ is influenced not just by the classical Lyapunov exponent. 

To demonstrate the  exponential growth in Eq.(\ref{eq:approtoc}) numerically, we turn to the continuous time evolution in a two-site Bose-Hubbard model\cite{PhysRevA.93.023621,PhysRevE.101.010202,WuLiu2006}.  There are two reasons for this choice: (1) this model is
simple enough for reliable numerical study; (2) the time evolution 
of the kicked rotor is discretized and  not convenient for analyzing short-time behavior.  This Bose-Hubbard model is also studied 
in Ref.~\cite{Xu2020} as the LMG model. Its Hamiltonian is given by (with $\hbar=1$)~\cite{WuLiu2006}
\begin{equation}
  \hat H = \frac 1 2 (\hat a_1^\dagger \hat a_0+\hat a_0^\dagger\hat a_1) - \frac {1} {N} (\hat a_1^\dagger \hat a_1-\hat a_0^\dagger \hat a_0)^2 ,
\end{equation}
where $\hat{a}_{0,1},\hat{a}_{0,1}^\dagger$ are annihilation and creation operators, $N$ is the total number of bosons in the system. For this system, the  effective Planck constant is $1/N$. When $N\rightarrow \infty$, the system becomes classical in the sense of mean-field approximation\cite{RevModPhys.54.407,Frohlich2007}. The system can be described by a single-particle Hamiltonian system of $H(q,p)=\sqrt{1/4-p^2}\cos q - 4 p^2$. There is a saddle point at $x^* = (q^*,p^*)=(\pi,0)$ and the Lyapunov exponent of it is $\lambda_C=\sqrt{3}$ 
(see Appendix \ref{mftheory} for details). 

\begin{figure}[t]
  \includegraphics[width=\columnwidth]{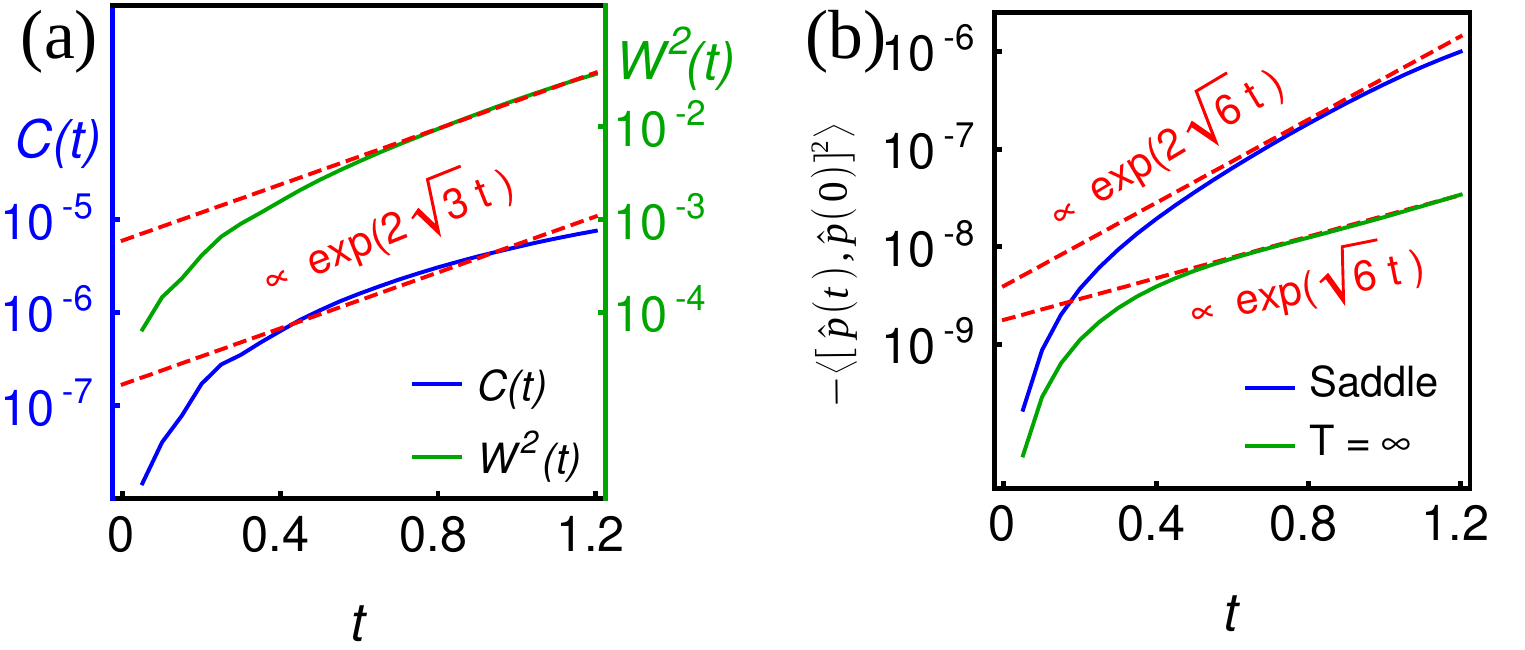}
  \caption{(color online) Numerical results of LMG model of $N=1681$ and $\hbar_{\text{eff}}\sim 1/N = 0.0005$. 
  (a) Our OTOC with macroscopic momentum $\hat P$ at saddle point (blue line). The green line is the width 
  of wavepacket along $P$-axis $W^2(t)$. Two red lines are both of exponent $2\sqrt{3}=\lambda_C$.
  (b) The OTOC with momentum $\hat p$. 
  The blue line is for its expectation value over $\ket{x^*}$, the Planck cell basis of classical saddle point;
  the green line is the usual expectation over the thermal state at  $T=\infty$. }
  \label{fig:lmgDyna}
\end{figure}

We focus on the quantal system of finite but large $N$. The Planck cell basis of discretized form is defined with the eigenstates of momentum operator $\hat p = (\hat a_0^\dagger \hat a_0 - \hat a_1^\dagger \hat a_1)/2N$~\cite{WuLiu2006,Zhao2019}(see Appendix \ref{pcdc}). 
In Fig. \ref{fig:lmgDyna}(a), the growth of OTOC of $C(t) = \bra{x^*}[\hat P(t),\hat P(0)]^2\ket{x^*}$ 
at the saddle point is shown. It is clear from the fitting that the  growth of $C(t)$ is exponential with the double of classical 
Lyapunov exponent of $2\sqrt{3}$, which agrees with our analysis in Eq. (\ref{eq:approtoc}).
As a comparison, we have also computed the OTOC in the form of $-\langle [\hat p(t), \hat p(0)]^2\rangle$, which was discussed in Ref.~\cite{Xu2020}. Our computation is done both at infinity temperature and on the Planck cell of saddle point. The results are shown in 
Fig. \ref{fig:lmgDyna}(a). Although we also see  exponential growth, however, both exponents are different from the classical Lyapunov exponent. 
We also plot the growth of the width $W^2(t)$ of the wavefunction along axis of $P$ in Fig.~\ref{fig:lmgDyna}(b). It is an approximation to Eq.~(\ref{eq:expect}) by $W^2(t) = \bra{x^*}\hat U^\dagger(\hat P(0)-0)^2\hat U\ket{x^*}\approx\sum_{z}(P_{\tg z}-P_{\tg x^*})^2|f(\tg z,x^*)|^2\sim C(t)$. We can find that its growth also fits the classical Lyapunov exponent well.

Note that we have replaced $Q$ with $P$ in Eq.~(\ref{eq:expect}) in the above computation for the convenience of comparison
to the results in Ref.~\cite{Xu2020}. Technically, one is allowed to replace $Q$ with $P$ or $P$ with $Q$ in Eq.~(\ref{eq:expect})
since  $\hat Q$ and $\hat P$ commute.

As indicated by the results in Fig. \ref{fig:lmgDyna}, the exponent of OTOC can be different from the  classical Lyapunov exponent. 
This difference has been noticed before~\cite{Rozenbaum2017, PhysRevB.98.134303}. Our numerical result shows that it is deeply related to the choice of operators. The OTOC with $\hat{p}$ at the saddle point (the blue line in Fig. \ref{fig:lmgDyna}(b))  grows different from classical Lyapunov exponent, while our OTOC with $\hat{P}$ does not.

A probable explanation for the inconsistency is that the classical limit for LMG model is different from systems with spatial degrees of freedom. There is no canonical commutation relation of $[\hat q, \hat p]=\ti \hbar$ in the finite-dimensional Hilbert space. The classical dynamics of LMG model is described as the transportation among Planck cell basis (the zeroth order in Eq. (\ref{eq:expan})). Since Eq. (\ref{eq:expect}) only relies on this property of quantum evolution, our OTOC and the analysis still work for LMG model. It is shown in Ref.~\cite{Cotler2018}  that under the Wigner-Weyl transformation, the operator $-[\hat q(t),\hat p(0)]^2$ leads to a function characterized by classical Lyapunov exponent in phase space in the classical limit. We illustrate the thermal OTOC of inverted harmonic oscillator\cite{Hashimoto2020} in Appendix \ref{biho}. In this system with intrinsic symplectic form, exponential growth of OTOC of $\hat q$ and $\hat Q$ shares the same exponent and agrees with the classical value.

{\it Conclusions} -- We have introduced a state-dependent form of OTOC, which serves as a microscope for resolving 
the fine structure in quantum dynamics. With analytical derivation, we have extracted and explained two sources 
of the growth of this OTOC. One controls its short-time behavior  while the other determines its long-time saturation. 
The early-time exponential growth of OTOC is caused by the former and related to the classical Lyapunov exponent. 
We are also able to show explicitly how the operator correlation in OTOC is related to wave packet spreading in Hilbert space.

\acknowledgements

This work is supported by the the National Key R\&D Program of China (Grants No.~2017YFA0303302, No.~2018YFA0305602),
National Natural Science Foundation of China (Grant No. 11921005), and
Shanghai Municipal Science and Technology Major Project (Grant No.2019SHZDZX01).

%

\clearpage
\appendix
\section{Basis of Quantum  Phase Space}\label{phbsproperty}
We focus on one dimensional system; the results can be generalized straightforwardly to higher dimensions. 
For a one dimensional classical system, its phase pace is two dimensional. In quantum statistical mechanics, 
one usually divides the phase space into Planck cells to obtain quantum phase space. 
In 1929, von Neumann proposed to construct a set of orthonormal basis by 
assigning each Planck cell a localized wave function~\cite{Neumann1929,vonNeumann2010}. 
This idea has been further developed with the help of the Wannier functions~\cite{Han2015,Fang2018,Jiang2017}.

One  basis for such a quantum phase space can be constructed with the following basis function
\begin{equation}
  \ket{Q,P} = \frac 1 {\sqrt{\Delta_q}} \int_Q^{Q+\Delta_q} \td q \ \ket{q} e^{\ti P q/\hbar}\,.
\end{equation}
Here $q$ is the position of a particle and $\ket{q}$ is the eigenstate of $\hat{q}$. $\Delta_q$ is one side of the Planck cell
and the other side is $\Delta_p$ such that $\Delta_q\Delta_p = 2\pi \hbar$. $Q,P$ are discretized position and momentum of
a given Planck cell. 

They are orthonormal:  $\bra{Q',P'}Q,P\rangle = \delta_{Q,Q'}\delta_{P,P'}$. Here is the proof. 
\begin{equation}
  \begin{aligned}
    & \ \ \ \bra{Q',P'}Q,P\rangle\\
    &= \frac 1 {\Delta_q} \int_{Q}^{Q+\Delta_q}\int_{Q'}^{Q'+\Delta_q}\td q\td q' \ \delta(q-q') e^{\ti P q/\hbar - \ti P'q'/\hbar}\\
    &=\frac 1 {\Delta_q} \delta_{Q,Q'} \int_Q^{Q+\Delta_q} \td q \ e^{\ti (P-P')q/\hbar} \\
    &= \frac {e^{\ti (P-P')Q/\hbar}} {\ti (P-P')/\hbar}\Big(e^{\ti(P-P')\Delta_q/\hbar} - 1\Big) \\
    &= 0 \ \ \ \text{ if } P\neq P'
  \end{aligned}
\end{equation}
where we have used the property that $P-P'\equiv 0 \mod \Delta_p$. That they are normalized can also be checked.

We can also construct the Planck cell basis with eigenstates of $\hat p$ as
\begin{equation}
  \ket{Q,P}_m = \frac 1 {\sqrt{\Delta_p}}\int_P^{P+\Delta_p} \td p \ \ket{p} e^{-\ti Qp/\hbar}\,.
\end{equation}
The sign difference in the exponent comes from the symplectic structure of classical mechanics. Only when we use such sign for $Qp$, we can get correct classical equation of motion from the quantum propagator at the limit of $\hbar\rightarrow 0$. 
This set of basis is consistent with the previous one, that is, 
\begin{equation}
  \lim_{\hbar\rightarrow 0} |\bra{Q',P'}Q,P\rangle_m|^2 = \delta_{Q,Q'}\delta_{P,P'}\,.
\end{equation}
One can check it as follows
\begin{equation}
  \begin{aligned}
    & \ \ \ \ \bra{Q',P'}Q,P\rangle_m \\
    &=\frac 1 {\sqrt{\Delta_q\Delta_p}}\int_{P}^{P+\Delta_p}\td p \int_{Q'}^{Q'+\Delta_q} \td q \ e^{-\ti Qp/\hbar -\ti P'q/\hbar} \bra{q}p\rangle \\
    &=\frac 1 {\sqrt{hV}}\int_0^{\Delta_p}\td p \int_0^{\Delta_q}\td q\ \exp\Big\{-\ti\big(Q(p+P) \\  
    & \ \ \ \  +P'(q+Q')-(p+P)(q+Q')\big)/\hbar\Big\} \\
    &=\frac {e^{\ti \alpha}} {\sqrt{hV}}\int_0^{\Delta_p}\td p \int_0^{\Delta_q}\td q\ e^{-\ti((Q-Q')p+(P'-P)q-qp)/\hbar} \\
  \end{aligned}
\end{equation}
At the limit of $\hbar\rightarrow 0$, the main contribution of the integral appears at the point of $Q=Q',P=P'$. 

\section{Mathematical Details for $C(t,x)$}\label{mdOTOC}
With  $x = (Q_x,P_x)$ and operators
\begin{equation}
  \hat Q = \sum_{x}\ket{x}Q_x\bra{x} \ ; \ \hat P = \sum_{x}\ket{x}P_x\bra{x},
\end{equation}
we have
\begin{eqnarray}
   && -C(t,x) = \bra{x}[\hat U^\dagger \hat Q \hat U,\hat P]^2\ket{x} \nonumber\\
    &=&\bra{x}\Big(\sum_{z,z',z''}\ket{z''}\bra{z'}(\bra{z''}\hat U^\dagger\ket{z}\bra{z}\hat U\ket{z'}Q_{z}P_{z'} \nonumber\\
    &&  - \bra{z''}\hat U^\dagger \ket{z}\bra{z}\hat U\ket{z'}Q_{z}P_{z''})\Big)^2\ket{x} \nonumber\\
    &=&-\sum_{z_1,z_2,z_3}\bra{x}\hat U^\dagger\ket{z_1}\bra{z_1}\hat U\ket{z_2}\bra{z_2}\hat U^\dagger\ket{z_3} \nonumber\\
    && \cdot \bra{z_3}\hat U\ket{x} Q_{z_1}Q_{z_3}(P_{z_2}-P_x)^2\,.
\end{eqnarray}
Since the propagator reads
\begin{equation}
  \bra{x'}\hat U\ket{x} = e^{\ti \phi(x)}\delta_{x',\tg x} + f(x',x) ,
\end{equation}
we can compute $C(t,x)$ by the orders of $f$. The zeroth order of $C$ is made up with four Kronecker symbols. 
Then the product leads to factor of $\delta_{z_2,x}$, this factor together with $(P_{z_2}-P_x)^2$ makes the zeroth order vanish. 
For the same reason, the first order of $C$ contains a product of three Kronecker symbols, $z_2$ will always be connected with $x$ and contributes a zero factor. Thus the leading term of $C(t,x)$ should be the second order of $f$. 
With the omission of two terms in which the product of Kronecker symbols connects $x$ and $z_2$, the nonzero terms of the second order are
\begin{equation}
  \begin{aligned}
    C^{(2)} &= \sum_z f^*(z,x)f(z,x) Q_z^2 (P_{\tg^{-1} z} - P_x)^2 \\
    +  \sum_z & f(\tg x,z)f^*(\tg x, z) Q_{\tg x} Q_{\tg x}(P_z-P_x)^2 \\
    + \sum_z & f(\tg x,z) f(\tg z,x) Q_{\tg x} Q_{\tg z} (P_z-P_x)^2 e^{-\ti \phi(x)-\ti\phi(z)}\\
    +  \sum_z & f^*(z,x)f^*(\tg x,\tg^{-1}z) Q_z Q_{\tg x} (P_{\tg^{-1}z}-P_{x})^2 \\
    \ \ \ \ \ \ \ \ &\cdot  e^{\ti \phi(\tg^{-1}z)+\ti \phi(x)} \\
    =\sum_z & (P_z-P_x)^2 \Big|Q_{\tg z}f(\tg z,x)e^{-\ti \phi(x)}\\
    &+Q_{\tg x}f^*(gx,z)e^{\ti \phi(z)}\Big|^2
  \end{aligned}
\end{equation}
With the unitarity of expansion of time evolution operator (up to the first order of $f$), we have
\begin{equation}
  \begin{aligned}
  \delta_{x,x'} &= \sum_z \bra{x'}\hat U\ket{z}\bra{z}\hat U^\dagger\ket{x}\\
   &= \sum_z (e^{\ti \phi(z)}\delta_{x',\tg z} + f(x',z))(e^{\ti \phi(z)}\delta_{x,\tg z} + f(x,z))^* \\
   &= \delta_{x,x'} + e^{\ti \phi(\tg^{-1} x')} f^*(x,\tg^{-1} x')\\
   & \ \ \ \ + e^{-\ti \phi(\tg^{-1} x)} f(x',\tg^{-1}x) + \cdots
 \end{aligned}
\end{equation}
This leads to an equality of
\begin{equation}
  e^{\ti \phi(x')} f^*(\tg x, x') + e^{-\ti \phi(x)} f(\tg x',x) = 0
\end{equation}
Thus, we have the second order of $C(t,x)$:
\begin{equation}
  C^{(2)} = \sum_z (P_z-P_x)^2 (Q_{\tg z}- Q_{\tg x})^2 |f(\tg z, x)|^2
\end{equation}
This is what we discussed in the main text.

\section{LMG Model and its Mean-Field Theory}\label{mftheory}
We consider a two-mode interacting Bose gas with the following 
 second quantized Hamiltonian 
\begin{equation}
  \hat H = \frac 1 2 (\hat a_1^\dagger \hat a_0+\hat a_0^\dagger \hat a_1)+\frac \xi {2N} \Big(\hat a_1^\dagger \hat a_1-\hat a_0^\dagger \hat a_0\Big)^2
\end{equation}
Its mean field theory can be obtained with  the coherent path integral 
\begin{equation}
  U(z_f^*,t_f;z_i,t_i) = \int \mathcal{D}z\mathcal{D}z^* e^{\ti S/\hbar}
\end{equation}
in which the action $S=\int_{t_i}^{t_f} \td t \ (\ti\hbar z^* \dot{z} - \bra{z}\hat H\ket{z})$ with
the coherent state $\ket{z}=e^{z_0\hat a_0^\dagger +z_1\hat a_1^\dagger}\ket{0}$.  $S/N$ should be a $O(1)$ quantity; 
we then rewrite the propagator as $U=\int e^{\ti (N/\hbar)\cdot (S/N)}$. Obviously the effective Planck constant is $\hbar/N$. 
 Then with the substitution $z\rightarrow x\sqrt{N}$ and letting $\hbar = 1$, the mean-field equation of motion can be obtained with the method of steepest gradient  (note the constraint of $|x_0|^2+|x_1|^2=1$):
\begin{widetext}
  \begin{equation}
  \ti\frac {\td} {\td t}\begin{bmatrix} x_0\\x_1\end{bmatrix} = \frac 1 {\sqrt{N}}\frac {\delta} {\delta z^*} \bra{z}\hat H\ket{z}=\begin{bmatrix}
  \xi(|x_0|^2-|x_1|^2)+\xi/2N & 1/2\\
  1/2 & -\xi(|x_0|^2-|x_1|^2)+\xi/2N
  \end{bmatrix}\begin{bmatrix} x_0\\x_1\end{bmatrix}
  \end{equation}
\end{widetext}
For a mean-field state $(x_0,x_1)$, its corresponding quantum state is 
\begin{equation}
  \ket{\Psi(x_0,x_1)} = \frac 1 {\sqrt{N!}}(x_0\hat a_0^\dagger +x_1\hat a_1^\dagger)^N\ket{0}\,.
\end{equation}
With the transformation of $p=\frac 1 2 (|x_0|^2-|x_1|^2); q=\arg x_1-\arg x_0$, one can find that 
this system is a Hamiltonian system with Hamiltonian (up to the order of $1/N$)
\begin{equation}
  H(q,p) = \sqrt{1/4-p^2}\cos q+ 2\xi p^2\,. \label{eq:clslmgenergy}
\end{equation}

The saddle point appears at $(q^*,p^*)=(\pi,0)$ when $\xi=-2$, and the classical Lyapunov exponent can be determined by  linearizing the canonical equation $\dot{q} = \partial_p H, \dot{p}=-\partial_q H$ near the saddle point,
\begin{equation}
  \begin{bmatrix} \dot{\delta q} \\ \dot{\delta p}\end{bmatrix} = \begin{bmatrix}
  \frac {\partial^2 H} {\partial q\partial p} & \frac {\partial H} {\partial p^2} \\
  -\frac {\partial^2 H} {\partial q^2} & -\frac {\partial^2 H} {\partial q \partial p}
  \end{bmatrix}\begin{bmatrix} \delta q\\ \delta p\end{bmatrix}
\end{equation}
The matrix is
\begin{equation}
  \begin{bmatrix}
  \frac {\partial^2 H} {\partial q\partial p} & \frac {\partial H} {\partial p^2} \\
  -\frac {\partial^2 H} {\partial q^2} & -\frac {\partial^2 H} {\partial q \partial p}
\end{bmatrix}\Bigg|_{(q,p)=(\pi,0)} = \begin{bmatrix}
0 & 4\xi+2 \\
-1/2 & 0
\end{bmatrix}
\end{equation}
which has the spectrum of $\pm \sqrt{3}$ when $\xi=-2$, that is our Lyapunov exponent for this saddle point.

The operator corresponding to $p$ is 
\begin{equation}
\hat p = (\hat a_0^\dagger \hat a_0 - \hat a_1^\dagger \hat a_1)/2N\,.
\end{equation}
Its eigenstates of  is  $\ket{s,N-s}=\frac 1 {\sqrt{s!(N-s)!}}\hat a_0^{\dagger s}\hat a_1^{\dagger N-s}\ket{0}$ with 
\begin{equation}
  \hat p\ket{s,N-s} = \frac {N-2s} {2N} \ket{s,N-s}\,.
\end{equation}

\subsection{Phase Space Basis for LMG Model}\label{pcdc}

For systems like the LMG model with finite dimension of Hilbert space, the definition in Eq.~\ref{eq:qpbasis} needs to be modified. 
According to Appendix.~\ref{phbsproperty}, we use eigenstates of $\hat p$. 
The phase space is divided into  $L\times L$ cells. The dimension of Hilbert space is $D = N+1 = L^2$. Letting $L = 2m+1$ be an odd number, the eigenstates of $\hat p$ is $\{\ket{p = n/N}\}_{n = -2m^2-2m}^{2m^2+2m}$. Then the basis function for the quantum phase space basis reads
\begin{equation}
  \ket{Q,P} = \frac 1 {\sqrt{L}} \sum_{n = -m+ PN}^{m+PN} \ket{p = \frac n N} e^{-\ti N Q p} 
\end{equation}
with lattice points $(Q,P) = (2\pi n_1 /L, L n_2 / N)$ in which $n_1 \in \{0,\cdots,L-1\}$ and $n_2 \in \{-m, -m+1,\cdots,m\}$. These states are orthonormal
\begin{equation}
  \begin{aligned}
  \bra{Q',P'} Q,P\rangle &=\frac 1 L \sum_{n,n'} \delta_{n,n'} \exp\Big(-\ti N(\frac {Qn} N - \frac {Q'n'} N)\Big) \\
  &=\frac 1 L \delta_{P,P'}\sum_{n=-m+PN}^{m+PN} e^{-\ti n(Q-Q')} \\
  &= \delta_{P,P'} \delta_{Q,Q'}
  \end{aligned}
\end{equation}
Since the effective Planck constant is proportional to $1/N$ for this system, there is also the classical limit like Eq.~\ref{eq:expan}.

\subsection{Classical Limit of LMG Model}\label{cmlmg}
\begin{figure}[t]
  \includegraphics[width = \columnwidth]{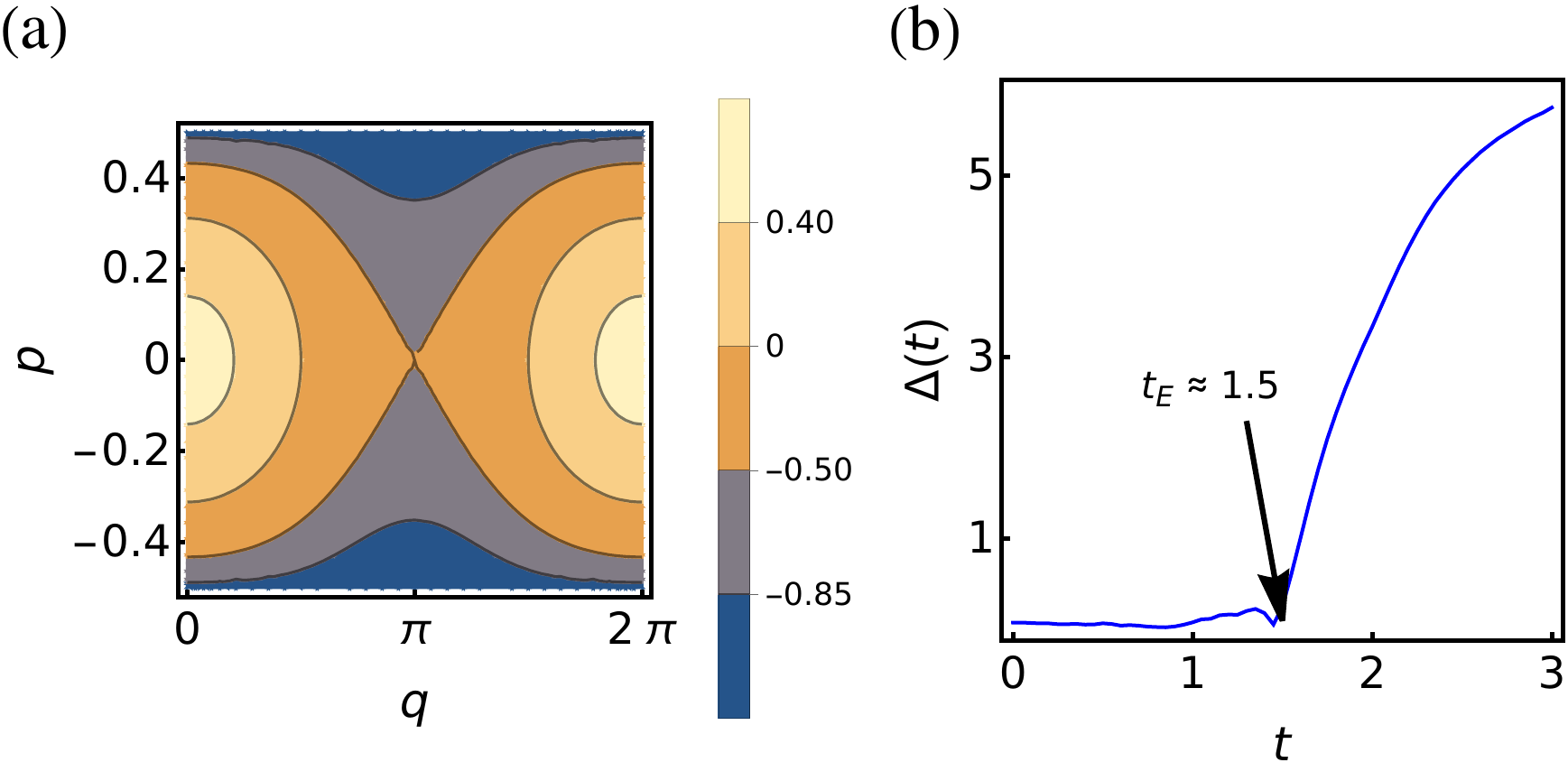}
  \caption{(color online) The LMG model in the classical limit. (a) The energy contours  for the classical Hamiltonian in the phase space. (b) The difference between the time evolution for quantum system at $N=1681$ and the classical one. $\Delta(t)$ measures the $L_2$ distance between quantum expectation value of macroscopic operators and the classical trajectory at time $t$.}
  \label{fig:clslmg}
  \end{figure}

The energy contour plot of classical Hamiltonian of LMG model (\ref{eq:clslmgenergy}) is shown in Fig.~\ref{fig:clslmg}(a). 
The saddle point is at $(q^*,p^*)=(\pi, 0)$. We consider an initial state at $\ket{Q=\pi, P=0}$. 
We let it evolve according to the second quantized Hamiltonian and the mean-field Hamiltonian, respectively.  
To compare them, we compute the expectation values
\begin{equation}
\langle \hat A(t) \rangle = \bra{Q=\pi, P=0}e^{\ti \hat H t}\hat Ae^{-\ti \hat H t}\ket{Q=\pi, P=0}\,,
\end{equation}
where $\hat A$ is either $\hat Q$ or $\hat P$. We finally compute  the difference 
\begin{equation}
  \Delta(t) = \sqrt{(q(t) - \langle \hat Q(t)\rangle)^2 + (p(t) - \langle \hat P(t)\rangle)^2} ,
\end{equation}
where  $(q(t), p(t))$ is the classical trajectory beginning with $(q(0) = q^* - 2\pi / L, p(0) = p^*)$. This little shift is necessary 
because the saddle point is a fixed point for the classical dynamics. The time evolution of $\Delta(t)$ is shown in Fig.~\ref{fig:clslmg}(b), 
where we find that before the finite time $t_E \approx 1.5$ it is close to zero and it grows rapidly after that. 
In this sense, $t_E$ is the Ehrenfest time for this saddle point. In the main text, our discussion of  the exponential growth of the 
OTOC is before this characteristic time.

\section{Inverted Harmonic Oscillator}\label{biho}

\begin{center}
\begin{figure}[h]
  \center
  \includegraphics[width = 0.9\columnwidth]{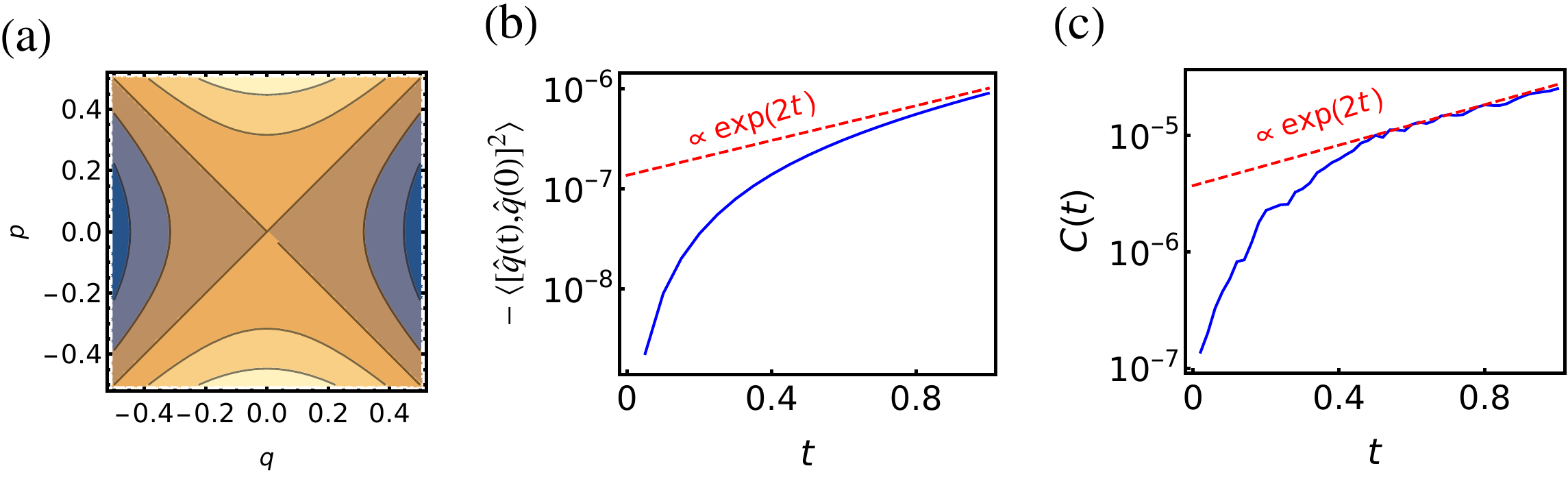}
  \caption{ (color online)
   Inverted harmonic oscillator with $\hbar=0.0005$. (a) The classical energy contours in phase space. A saddle point is at $(q^*, p^*) = (0,0)$ with the classical Lyapunov exponent $\lambda_C = 1$. (b) The growth of OTOC of $-\text{Tr} \hat \rho_T [\hat q(t),\hat q(0)]^2$ with $\hat \rho_T = e^{-\hat H/T} / \text{Tr}e^{-\hat H/T}$ and $T = 0.1$. 
    (c) The growth of OTOC of $C(t) = -\bra{Q=0,P=0}[\hat Q(t),\hat Q(0)]\ket{Q=0,P=0}$. The exponential growth of these two OTOCs are both twice of the classical Lyapunov exponent. 
  }
  \label{fig:bihoDyna}
\end{figure}
\end{center}

The Hamiltonian of an inverted harmonic oscillator reads
\begin{equation}
  H(q,p) = \frac 1 2 (p^2 - q^2)
\end{equation}
By the same analysis above, it has a saddle point at $(q^*, p^*) = (0, 0)$ with energy $0$, whose Lyapunov exponent is $\lambda_C = 1$. 
In quantum regime, the dynamics of wave function obeys the \sdg~equation:
\begin{equation}
  \ti \hbar \frac {\partial \psi} {\partial t} = -\frac {\hbar^2} {2} \frac {\partial^2\psi} {\partial q^2} + V(q)\psi(q)
\end{equation}
Here we add a hard wall to the potential so that $q$ is confined by 
\begin{equation}
V(q) = \begin{cases}
  - q^2 / 2 & q \in [-1/2, 1/2]\\
  \infty & \text{otherwise}
\end{cases}. 
\end{equation}
This modification will not change the dynamics near the saddle point. Different from the LMG model, inverted harmonic oscillator has a infinite dimensional Hilbert space. Numerically, we need a cutoff for momentum, which is related to the coordinate resolution. Here, we choose $\hbar = 0.0005$, with the momentum cutoff at $|p| \leq 1 = p_{\text{max}}$, the coordinate is discretized by interval $\delta x \approx 0.0003$. The size of Planck cells is $\Delta_q = \sqrt{2\pi \hbar} \approx 0.056$. This finite shift along $q$ diverges exponentially and in order to the numerical accuracy of operator and correlation function, we can only consider the time evolution at the beginning. In Fig. \ref{fig:bihoDyna}(b), (c), we illustrate the growth of OTOC of $-\langle [\hat q(t),\hat q(0)]^2\rangle_T$ and $-\bra{Q=0,P=0}[\hat Q(t),\hat Q(0)]^2\ket{Q=0,P=0}$. We choose $T=0.1$ so that there are $80\%$ cumulative probability of the states of energy lower than $0.05$, i.e., most contribution are made by the region near the saddle point. It is clear that the growth of these two OTOCs are both exponential with the classical Lyapunov exponent. 

\onecolumngrid

\twocolumngrid

\end{document}